\documentclass[useAMS,usenatbib,twocolumn]{mn2e}
\usepackage{natbib}
\usepackage{graphicx}
\usepackage{amssymb,latexsym,amsopn,amsbsy}
\usepackage{pifont}% http://ctan.org/pkg/pifont
\newcommand{\cmark}{\ding{51}}%
\newcommand{\xmark}{\ding{55}}%
{\tiny }\usepackage{epsfig}
\usepackage[T1]{fontenc}
\usepackage{mathtools}

\usepackage[plainpages=false, colorlinks=true, anchorcolor=blue, linkcolor=blue, citecolor=blue, bookmarks=false]{hyperref}

\usepackage{color}

\newcommand{\be}{\begin{equation}}
\newcommand{\ee}{\end{equation}}
\newcommand{\bear}{\begin{eqnarray}}
\newcommand{\eear}{\end{eqnarray}}

\newcommand{\xp}{x_{\rm p}}
\newcommand{\mut}{\tilde{\mu}}

\newcommand{\cT}{{\cal T}}

\newcommand{\cA}{{\cal A}}

\newcommand{\cS}{{\cal S}}
\newcommand{\cM}{{\cal M}}
\newcommand{\rx}{{\rm x}}

\newcommand{\rn}{{\rm n}}
\newcommand{\rp}{{\rm p}}
\newcommand{\re}{{\rm e}}
\newcommand{\rpe}{{\rm pe}}
\newcommand{\rc}{{\rm c}}

\newcommand{\rs}{{\rm s}}

\newcommand{\rT}{{\rm T}}

\newcommand{\rL}{{\rm L}}

\newcommand{\rP}{{\rm P}}
\newcommand{\rH}{{\rm H}}
\newcommand{\rF}{{\rm F}}
\newcommand{\rEM}{{\rm EM}}
\newcommand{\bnabla}{\boldsymbol{\nabla}}
\newcommand{\bB}{\mathbf{B}}
\newcommand{\bF}{\mathbf{F}}

\title[The freedom to choose neutron star MHD equilibria]{The freedom to choose neutron star magnetic field equilibria}

\author[Glampedakis \& Lasky]
{
Kostas Glampedakis$^{1,2}$\thanks{E-mail: kostas@um.es}
and Paul D. Lasky$^{3}$\thanks{E-mail: paul.lasky@monash.edu}, 
\\
$^{1}$Departamento de F\'isica, Universidad de Murcia, Murcia, E-30100, Spain\\
$^{2}$Theoretical Astrophysics, University of T\"ubingen, Auf der Morgenstelle 10, T\"ubingen, D-72076, Germany\\
$^{3}$Monash Centre for Astrophysics, School of Physics and Astronomy, Monash University, VIC 3800, Australia
}

\begin{document}

\label{firstpage}

\maketitle

\begin{abstract}

Our ability to interpret and glean useful information from the large body of observations of strongly magnetised neutron stars 
rests largely on our theoretical understanding of magnetic field equilibria. We answer the following question: is one free to arbitrarily
prescribe magnetic equilibria such that fluid degrees of freedom can balance the equilibrium equations?
We examine this question for various models for neutron star matter; from the simplest single-fluid barotrope to more realistic 
non-barotropic multifluid models with superfluid/superconducting components, muons and entropy.  
We do this for both axi- and non-axisymmetric equilibria, and in Newtonian gravity and general relativity.  
We show that, in axisymmetry, the most realistic model allows complete freedom in choosing a magnetic field equilibrium 
whereas non-axisymmetric equilibria are never completely arbitrary.  

\end{abstract}


%%%%%%%%%%%%%%%%%%%%%%%%%%%%%%%%%%%%%%%%%%%%%%%%%%%%%%%%%%%%%

\section{Introduction}
\label{sec:intro}
The structure of magnetic fields in the interior of neutron stars has undergone intense scrutiny in recent years.
The endeavour has been mostly motivated by X-ray observations of strongly magnetised neutron stars.  For example, bursts and giant flares 
occurring in magnetars are commonly believed to be powered by these objects' super-strong 
magnetic fields \citep[see][for reviews]{woods06,mereghetti15}.  Other isolated neutron 
stars---colloquially known by the sobriquet `magnificent seven'---are kept warm by their evolving magnetic fields and emit intense 
thermal radiation \citep{haberl07}. 
In fact, the resulting theory-observations synergy appears to converge to a notion of a magnetic field-dominated
evolutionary link between various sub-families of the neutron star population \citep{kaspi10,vigano13}.
   
For the neutron star theorist wishing to delve into the physics of neutron star magnetic fields, a reasonable
starting point is determining the nature of magnetic field equilibria.    
There is a significant corpus of recent work on that topic, both analytical 
~\citep[e.g.,][]{haskell08, reisenegger09,ciolfi09,mastrano11,glampedakis12a,lasky13b,ciolfi13} 
and numerical~\citep[e.g.,][]{braithwaite07,braithwaite09,  lasky11,kiuchi11,ciolfi12,lander12,lander12b,fujisawa13,palapanidis15,bucciantini15}. 
One important result that has emerged is that magnetohydrodynamic (MHD) equilibria are greatly influenced
by the properties of matter or, in other words, by the equation of state (EOS). The key physics factor is that of \textit{barotropic} 
versus \textit{non-barotropic} matter, the latter allowing a much larger family of magnetic field equilibria \citep{reisenegger09}.
A closely related factor is that of the imposed \textit{symmetries} in the system: for practical reasons 
most existing work assumes axisymmetric (2-D) magnetic equilibria; very little is known of non-axisymmetric (3-D) equilibria. 

%%%%%%%%
\begin{table*}
\begin{minipage}{135mm}
	\caption{Overview of results.  For each type of equation of state, symmetry of the magnetic field and 
	type of gravity (i.e., Newtonian/general relativistic), we calculate whether an arbitrary magnetic field 
	can be balanced by the degrees of freedom in the fluid, and hence whether one is arbitrarily free to 
	prescribe a magnetic field.  The final column of the table lists the section of the paper in which the relevant calculations can 
	be found.}\label{tab:overview}
	\begin{tabular}{rcccc}
	\hline
	Equation of state & Symmetry & Gravity & Free to prescribe? & Section\\
	\hline\hline
	%\multicolumn{3}{c}{Newtonian Gravity}\\
	%\hline
	Single fluid, barotropic & axi- \& non-axisymmetric & Newtonian & \xmark & \ref{sec:baro}\\
	Single fluid, non-barotropic & axisymmetric & Newtonian & \cmark & \ref{sec:axi}\\
	Single fluid, non-barotropic & non-axisymmetric & Newtonian & \xmark & \ref{sec:nonaxi}\\
	Single fluid, elastic crust & axi- \& non-axisymmetric & Newtonian & \cmark & \ref{sec:crust}\\
	Multifluid, cold $npe$ matter & axi- \& non-axisymmetric & Newtonian & \xmark & \ref{sec:coldsf} \\
	Multifluid, hot $npe\mu$ matter & axisymmetric & Newtonian & \cmark & \ref{sec:hotsfaxi}\\
	Multifluid, hot $npe\mu$ matter & non-axisymmetric & Newtonian & \xmark & \ref{sec:hotsfnonaxi}\\
	\hline
	%\multicolumn{3}{c}{General Relativity}\\
	%\hline
	Single fluid, barotropic & axisymmetric & general relativistic & \xmark & \ref{sec:baroGS}\\
	Single fluid, non-barotropic & axisymmetric & general relativistic & \cmark & \ref{sec:noGS}\\
	\hline 
	\end{tabular}
\end{minipage}
\end{table*}
%%%%%%%%%

This paper addresses a more `global' issue: \textit{to what extent MHD equilibria in neutron stars are arbitrary for 
given EOSs, with or without axisymmetry.} 
Here, `arbitrary' means that a magnetic field $\bB$ can be freely prescribed (assuming it obeys $\bnabla \cdot \bB=0$), 
knowing that the available fluid degrees of freedom can satisfy the equilibrium equations.

In order to address this crucial issue we start from the simplest case of a single-fluid barotropic model and 
subsequently consider more realistic neutron star models with stratification, superfluidity, departure from 
chemical equilibrium, muons and entropy. For each case we study both axisymmetric and non-axisymmetric 
equilibria. In some cases we find that the magnetic field {\it cannot} be freely prescribed, and is required 
to solve a Grad-Shafranov differential equation (i.e. the MHD equilibrium equation for the magnetic scalar 
`stream functions' and the fluid parameters of the non-magnetic star) while in other cases the field
can be completely `user-specified'. We also find that non-axisymmetry entails additional constraints for 
the magnetic field. A summary of our main results can be found below in Section~\ref{sec:overview}.

The paper is set out as follows: in \S\ref{singlefluid} we develop the general formalism for a single 
fluid star, applying this specifically to barotropic and stratified matter in \S\ref{sec:baro} and 
\S\ref{sec:stratified}, respectively. We apply specific examples of axi- and non-axisymmetric magnetic field 
equilibria in \S\ref{sec:axi} and \S\ref{sec:nonaxi}, respectively. In \S\ref{sec:crust} we digress 
and discuss magnetic field equilibrium in neutron star crusts.
In \S\ref{sec:2fluid} we generalise the formalism to multifluid neutron stars with superfluid/superconducting 
components. We first treat the case of cold $npe$ matter and subsequently move to the most realistic model 
considered in this paper, accounting for the presence of muons and a finite temperature.
Finally, in \S\ref{sec:GR} we consider  full general relativistic (GR) MHD equilibria with both barotropic and
non-barotropic EOS.

%%%%%%%%%%%%%%%%%%%%
\subsection{Overview of results}
\label{sec:overview}

This section provides a compact summary of our main results, having in mind the `fast-track' reader 
who may not have a penchant for detailed calculations and numerous equations. Our findings are 
shown in Table~\ref{tab:overview}.

We can categorise our results in the following way:

\begin{enumerate}

\item\label{list:one} assuming axisymmetry, we first recover the well-known results for single-fluid barotropic stars 
(the field is required to solve a Grad-Shafranov equation) and single-fluid non-barotropic 
stars (the field can be freely prescribed).  Moving to multifluid neutron stars (which are always 
non-barotropic) with cold $npe$ matter, we find that the system  effectively behaves as a barotrope with
an accompanying Grad-Shafranov equation. The addition of muons and entropy in the previous
model (hot $npe\mu$ matter) restores the complete freedom in prescribing an MHD
equilibrium. 

\item\label{list:two} non-axisymmetric equilibria in the core are \textit{never} completely arbitrary, even when the magnetic field
is not required to solve a Grad-Shafranov equation. This is due to the presence of an additional
constraint equation between the azimuthal and non-azimuthal magnetic force components.

\item the transition from Newtonian MHD to GRMHD does not produce any qualitative changes and the 
conclusions of \ref{list:one} and \ref{list:two} remain valid (this why we show just two cases of GRMHD equilibria). 

\item The magnetic equilibrium in the crust can be freely specified provided we allow for a strained crust and
the associated elastic force. 
\end{enumerate}

%%%%%%%%%%%%%%%%%%%
\section{Single fluid stars}
\label{singlefluid}

The starting point for considering MHD equilibria in neutron stars is that of 
a `single-fluid' model. This term is actually a misnomer because we actually account 
for the presence of various constituents (such as neutrons, protons and electrons). It is, 
however, assumed that these particles (modulo the much less massive electrons) 
always move in unison (on hydrodynamical scales) as a result of their mutual collisions.

Assuming for simplicity a non-rotating static system (adding rotation is trivial as it amounts to
adding a centrifugal term in the gravitational potential), the single fluid MHD equilibrium is 
described by the Euler equation
\be
\bnabla p+\rho\bnabla\Phi={\bf F}_\rL,
\ee
where $p$, $\rho$, and $\Phi$ are respectively the pressure, density and gravitational potential, and ${\bf F}_\rL$ is the 
Lorentz force given by
\begin{equation}
	{\bf F}_\rL=\frac{1}{4\pi}\left(\bnabla\times{\bf B}\right)\times{\bf B}.
\end{equation}

In this paper, we treat the magnetic force as a perturbation on some spherical background 
(an approximation well justified for all astrophysical magnetic fields), 
such that 
\begin{align}
	& \bnabla p+\rho\bnabla\Phi=0,
	\label{background}
\\
%\notag \\
	& \bnabla\delta p+\delta\rho\bnabla\Phi+\rho\bnabla\delta\Phi={\bf F}_\rL.
	\label{firstorder}
\end{align}
Here, Eqn.~(\ref{background}) is the background hydrostatic equilibrium, and Eqn.~(\ref{firstorder}) is the leading-order 
perturbation equation, where a $\delta$ denotes perturbed quantities and second-order terms have been neglected.
Combining (\ref{background}) and (\ref{firstorder}) implies
\be
	\bnabla\left(\delta h+\delta\Phi\right)+\frac{1}{\rho}\left(\delta p\bnabla\rho-\delta\rho\bnabla p\right)={\bf F}_\rL,
	\label{euler0}
\ee
where $\delta h\equiv\delta p/\rho$ is the enthalpy perturbation.

The possible geometry of a magnetic field depends on the matter content of the star through the EOS.  
In general the latter will have the functional form $p=p(\rho,\,x)$, where $x$ is typically identified with the neutron star composition 
(i.e., the proton fraction, $x_\rp$).  One typically defines a pair of adiabatic indices according to
\begin{align}
	\bnabla p &=\frac{p}{\rho}\Gamma_0\bnabla\rho,\label{Gamma0}
\\
	\Gamma_1 &=\frac{\rho}{p}\left.\frac{\partial p}{\partial\rho}\right|_x.\label{Gamma1}
\end{align}
Expanding the left hand side of Eqn.~(\ref{Gamma0}) and substituting Eqn.~(\ref{Gamma1}) leads to
\be
	\frac{p}{\rho}\left(\Gamma_0-\Gamma_1\right)\bnabla\rho=\frac{\partial p}{\partial x}\bnabla x.
\ee
Importantly, this equation implies that $\Gamma_0=\Gamma_1$ if and only if there is no stratification in the star.  

After some straightforward manipulations we can write the Euler equation (\ref{euler0}) in the
equivalent form,
\be
	\bnabla\left(\delta h+\delta\Phi\right)+\Lambda_\rho \bnabla\rho=\frac{1}{\rho}{\bf F}_\rL,
	\label{MHDEquil}
\ee
where 
\be
	\Lambda_\rho\equiv\frac{p}{\rho^3}\left[\left(\Gamma_1-\Gamma_0\right)\delta\rho
	 +\frac{\rho}{p}\frac{\partial p}{\partial x}\delta x\right].
\ee
We can rewrite $\Lambda_\rho$ without $\delta x$ appearing explicitly. From the EOS we have
\be
\delta h = \frac{p \Gamma_1}{\rho^2} \delta \rho + \frac{\partial p}{\partial x} \frac{\delta x}{\rho},
\label{dh_eq}
\ee
implying
\be
 \Lambda_\rho = \frac{1}{\rho} \left ( \, \delta h - \frac{p \Gamma_0}{\rho^2} \delta \rho  \, \right ). 
\label{Lrho2}
\ee
In order to proceed further we need to make a choice for the EOS. 
In the following sections we consider separately the cases of barotropic and non-barotropic matter.

%%%%%%%%%%%%%%%%%%%%%%%%%%%%%%%%%%%%%%%%%%%%%

\subsection{Barotropic matter}
\label{sec:baro}
We begin by considering barotropic matter in which the EOS is simply $p=p(\rho)$, which 
implies no stratification. In this case, $\Lambda_\rho=0$ and taking the curl of 
Eqn.~(\ref{MHDEquil}) results in
\be
\bnabla \times \left  ( \frac{1}{\rho} \bF_\rL  \right ) = 0,
\label{curl1}
\ee
which is an equation involving only the magnetic-field degrees of freedom and the \textit{background} density. 

If in addition the system is assumed to be \textit{axisymmetric}, we can decompose the field into poloidal and
toroidal components, $\mathbf{B} = \mathbf{B}_\rP + \mathbf{B}_\rT$, where
\be
 \mathbf{B}_\rP  = \bnabla \Psi  \times \bnabla \varphi, \quad
\mathbf{B}_\rT = T \bnabla \varphi,
\label{BP+T}
\ee
and the two degrees of freedom are represented by the scalar stream functions $\Psi (r,\theta)$ and $T(r,\theta)$ 
(here we use standard spherical coordinates).  
In axisymmetry, the azimuthal component of the Lorentz force must vanish, $F_\rL^{\varphi}=0$, and one can easily
show that $\bnabla \Psi \times \bnabla T =0 \Rightarrow T=T(\Psi)$. 
The full Lorentz force can be written as
\be
\bF_\rL = - \frac{1}{4\pi\varpi^2} \left (\,  \Delta_* \Psi + T T^\prime \, \right ) \bnabla \Psi  \equiv  \cA \bnabla \Psi,
\ee
where a prime denotes a derivative with respect to the argument, $T^\prime = dT/d\Psi$, $\varpi=r\sin\theta$, and
\be
\Delta_* \Psi = \nabla^2 \Psi - \frac{2}{r} \left [\,  \mathbf{\hat{r}} \cdot \bnabla \Psi  
+ \cot\theta ( \boldsymbol{\hat{\theta}} \cdot \bnabla \Psi )  \, \right ],
\ee
where $\mathbf{\hat{r}}$ and $\boldsymbol{\hat{\theta}}$ are unit vectors.

The so-called Grad-Shafranov equation can be derived from Eqn.~(\ref{curl1})
\be
\frac{\cA}{\rho} = M(\Psi) ~ \Rightarrow ~
\Delta_* \Psi + T T^\prime = 4\pi \varpi^2 \rho M,
\label{GradS}
\ee
where $M(\Psi)$ is an arbitrary function. 

Equation~(\ref{GradS}) implies that axisymmetric magnetic field equilibria in barotropic stars come from 
a restricted class of solutions. That is, a single equation governs both the poloidal and toroidal component, 
implying one is not free to arbitrarily choose both components independently.

In general, once $\mathbf{B}$ has been obtained from Eqn.~(\ref{GradS}), the perturbed fluid's degrees of 
freedom can be found from Eqn.~(\ref{MHDEquil}) with $\Lambda_\rho=0$.  We have,
\be
\bnabla ( \delta h + \delta \Phi) =  M  \bnabla \Psi.
\ee
This is solved by,
\be
\delta h + \delta \Phi = g (\Psi) = \int d\Psi M(\Psi).
\ee
The system is closed with the perturbed Poisson equation $\nabla^2 \delta \Phi  = 4\pi G \delta \rho$.

Next, we consider 3-D \textit{non-axisymmetric} magnetic field equilibria in barotropic matter. 
Although perhaps not so well known in the literature, it is still possible to parametrise $\bB$
in terms of a pair of scalar functions. These are the so-called Euler potentials 
$\{\alpha(r,\theta,\varphi), \beta(r,\theta,\varphi)\}$ and are defined\footnote{We note that the Euler potentials are not 
uniquely defined. 
For instance, (\ref{EP1}) is invariant with respect to the gauge transformation $\beta \to \beta + g(\alpha)$, 
where $g$ is arbitrary. } as~\citep[e.g. see][]{yahalom08}:
\be
\bB = \bnabla \alpha \times \bnabla \beta,
\label{EP1}
\ee
and as a consequence $\bnabla \cdot \bB =0$ is trivially satisfied.

In terms of the Euler potentials, the electric current  and the Lorentz force become,
\begin{align}
\mathbf{J} &=  \frac{c}{4\pi} \left [\nabla^2 \beta \bnabla \alpha -   \nabla^2 \alpha \bnabla \beta 
+  ( \bnabla \beta \cdot \bnabla ) \bnabla \alpha -  ( \bnabla \alpha \cdot \bnabla ) \bnabla \beta \right ],
\\
&\bF_\rL = h_{\alpha\beta} \bnabla \alpha + h_{\beta \alpha} \bnabla \beta,
\label{FL3D}
\end{align}
where
\begin{align}
h_{\alpha\beta} &\equiv  \frac{1}{4\pi} \mathbf{J} \cdot \nabla \beta 
\nonumber \\
& =\frac{1}{4\pi} \left [\, \nabla^2 \beta ( \bnabla \alpha \cdot \bnabla \beta) 
-\nabla^2 \alpha ( \bnabla \beta \cdot \bnabla \beta)
\right.
\nonumber \\
& \left. + \bnabla \beta \cdot ( \bnabla \beta \cdot \bnabla ) \bnabla \alpha 
- \bnabla \beta \cdot ( \bnabla \alpha \cdot \bnabla ) \bnabla \beta \, \right ].
\end{align}
Based on these results we attempt to derive a ``Grad-Shafranov'' equation for a general 
non-axisymmetric equilibrium. From (\ref{curl1}) we have,
\be
 \bnabla \left ( \frac{h_{\alpha\beta}}{\rho} \right ) \times \bnabla \alpha
+ \bnabla \left ( \frac{h_{\beta\alpha}}{\rho} \right ) \times \bnabla \beta = 0. 
\label{3DGSeq}
\ee
This is solved by
\be
h_{\alpha\beta} = \rho \cM (\alpha), \qquad h_{\beta\alpha} = \rho \tilde{\cM}(\beta),
\label{3DGS}
\ee
where $\cM, \tilde{\cM}$ are arbitrary functions (not necessarily different).
Equations~(\ref{3DGS}) can be viewed as the non-axisymmetric generalisation of the
2-D Grad-Shafranov equation (\ref{GradS}). 

To conclude this section, it is interesting to investigate the connection between the Euler potentials $\alpha,\beta$ 
and the stream functions $\Psi,T$ in axisymmetry. We have,
\be
\bB = \bnabla \Psi \times \bnabla \varphi + T \bnabla \varphi = \bnabla \alpha + \bnabla \beta.
\ee
The inner and cross product with $\bnabla \varphi$ leads to, respectively,
\begin{align}
T &
=\varpi ( \hat{\boldsymbol{\varphi}} \times \bnabla \alpha ) \cdot \bnabla \beta,
\\
\bnabla \Psi &= \partial_\varphi \beta \bnabla \alpha - \partial_\varphi \alpha \bnabla \beta.
\end{align}
At this point we can choose either $ \partial_\varphi \alpha=0$ or $\partial_\varphi \beta=0$.
Opting for the former,
\be
\bnabla \Psi \times \bnabla \alpha = 0 ~ \Rightarrow ~ \Psi =\Psi (\alpha),
\label{Psi_EP}
\ee
which shows that the $\alpha$ and $\Psi$ surfaces coincide. 
Also,
\be
\partial_\varphi \beta = \Psi^\prime ~ \Rightarrow ~ 
\beta = \Psi^\prime \varphi + K(r,\theta),
\ee
where $K$ is an integration `constant' and
\be
T=\varpi ( \hat{\boldsymbol{\varphi}} \times \bnabla \alpha ) \cdot \bnabla K.
\label{T_EP}
\ee
Equations~(\ref{Psi_EP}) and (\ref{T_EP}) are the desired relations between 
the Euler potentials and the stream functions in axisymmetry.

%%%%%%%%%%%%%%%%%%%%%%%%%%%%%%%%%%%%%

\subsection{Stratified matter}
\label{sec:stratified}
In contrast to the barotropic case, we show in this section that there is significantly more freedom in the stratified 
case, in the sense that any user-specified magnetic field can be in equilibrium.

For cold (i.e. zero temperature) neutron star matter, the appropriate second parameter in the EOS is the proton fraction, i.e. $x=\xp$. 
For simplicity, we adopt the Cowling approximation; note that this is not restrictive since $\delta \Phi$
comes with the perturbed Poisson equation. The relevant Euler equation is:
\be
\bnabla \delta h  + \Lambda_\rho  \bnabla \rho = \frac{1}{\rho} \bF_\rL,
\label{eulerv4}
\ee
with $\Lambda_\rho$ given by (\ref{Lrho2}).

%%%%%%%%%%%%%%%%%%%%

\subsubsection{Axisymmetric equilibria}
\label{sec:axi}
In axisymmetry, we again decompose the field into poloidal and toroidal components, and again have $T=T(\Psi)$ 
for the magnetic stream functions and $\bF_\rL = \cA \bnabla \Psi$. 
However, taking the curl of both sides of Eqn.~(\ref{eulerv4}) implies 
\begin{align}
& \bnabla \left ( \frac{\cA}{\rho} \right ) \times \bnabla \Psi = \bnabla \Lambda_\rho \times \bnabla \rho = 
\nonumber  \\
%\nonumber \\
&= \frac{\rho^\prime}{\rho} \left [\,   \mathbf{\hat{r}} \times \left \{  \frac{p \Gamma_0}{\rho^2} \bnabla \delta \rho 
-  \bnabla \delta h \right \}  \, \right ] \neq 0 
~\Rightarrow ~ \frac{\cA}{\rho} \neq M (\Psi),
\end{align}
where a prime in a background quantity denotes a radial
derivative.  The above equation means that there is no Grad-Shafranov equation.  
In other words, there is no single equation that governs both the toroidal {\it and} poloidal field components, implying these 
components can be specified independently from one another.  Faced with this result, we simply consider $\bB$ as 
given and investigate if the fluid parameters can be adjusted to ensure an MHD equilibrium.  

Unlike the barotropic case, the perturbations $\delta h$ and $\delta \rho$ are not required to be linearly dependent 
(the remaining function $\delta x_\rp$ follows trivially from the EOS relation once $\delta h$ and $\delta \rho$ are known, 
see Eqn.~(\ref{dh_eq})).

Expanding the Euler equation~(\ref{eulerv4}) in components:
\begin{align}
& \partial_r \delta p - \frac{p \Gamma_0 \rho^\prime}{\rho^2} \delta \rho = \cA \partial_r \Psi,\label{drp}
\\
& \partial_\theta \delta p =  \cA \partial_\theta \Psi.
\end{align}
Integrating the bottom equation
\be
\delta p = \zeta(r) +  \int d\theta  \cA (r,\theta) \partial_\theta \Psi (r, \theta)\label{delp}
\ee
where $\zeta(r)$ is an arbitrary  spherical `gauge'  function.   Finally, putting this back into Eqn.~(\ref{drp}) implies
\be
\delta \rho = \frac{\rho^2}{p \Gamma_0 \rho^\prime} \left [\,  \zeta^\prime   - \cA \partial_r \Psi 
+ \partial_r \left \{  \int d\theta  \cA (r,\theta) \partial_\theta \Psi (r, \theta) \right \} \, \right ].\label{delrho}
\ee
These equations show that there is sufficient freedom in the fluid variables to balance a given magnetic field.  
That is, for a given $\bB$, once one writes down the functional form for the poloidal and toroidal field, the 
Euler equations can be fully solved using Eqns.~(\ref{delp}) and (\ref{delrho}) for the perturbed density and pressure
such that the system is in MHD equilibrium.

%%%%%%%%%%%%%%%%%%%%%%%%%%%%%

\subsubsection{Non-axisymmetric equilibria}
\label{sec:nonaxi}

We again consider the equilibrium described by  the Euler equation (\ref{eulerv4}) but this time without the assumption of
axisymmetry -- this means that we now have three equations for the two unknown functions $\delta h$ and $\delta \rho$.
As a result, we expect to have a non-trivial relation between the magnetic force components. Let us see
how this works in detail.

Decomposing (\ref{eulerv4}) in its components:
\begin{align}
& \partial_r \delta h +  \frac{\rho^\prime}{\rho}  \left ( \, \delta h - \frac{p \Gamma_0}{\rho^2} \delta \rho  \, \right )  
=\frac{1}{\rho} F^r_\rL,
 \\
& \Rightarrow~  
\partial_r \delta p - \frac{p \Gamma_0 \rho^\prime}{\rho^2} \delta \rho = F_\rL^r,
\\
& \partial_\theta \delta h =  \frac{r}{\rho} F_\rL^\theta  ~ \Rightarrow ~ \partial_\theta \delta p 
= r F^\theta_\rL,
\label{eulerth1}
\\
& \partial_\varphi \delta h =  \frac{\varpi}{\rho} F_\rL^\varphi   ~ \Rightarrow ~ \partial_\varphi \delta p 
= \varpi F^\varphi_\rL.
\label{eulerph1}
\end{align}
The radial equation is the only one featuring $\delta \rho$ and therefore we solve it with respect to that parameter:
\be
\delta \rho = \frac{\rho^2}{p \Gamma_0 \rho^\prime} \left (\, \partial_r \delta p - F^r_\rL \, \right ).
\ee
We are left with two equations for $\delta p$. It is easy to combine them, eliminate $\delta p$, and arrive to 
a condition for the magnetic force components:
\be
\partial_\varphi F^\theta_\rL = \partial_\theta ( \sin\theta F^\varphi_\rL ).
\label{Fconstrain1}
\ee
We can subsequently use either of (\ref{eulerth1}) or (\ref{eulerph1}) to obtain $\delta p$. Using the former,
\be
\delta p  = \zeta (r,\varphi) + r \int d\theta F^\theta_\rL
\ee
where $\zeta$ is an arbitrary function. 

The condition (\ref{Fconstrain1}) implies that the non-axisymmetric system does not allow an arbitrarily specified $\bB$ 
field. It should be emphasized that this condition is valid for both stratified and barotropic matter (since both cases share
the same $\theta$ and $\varphi$ Euler components). Clearly, in the limit of an axisymmetric field, it
is  trivially satisfied.

We can further expand (\ref{Fconstrain1}) with the help of the Euler potentials $\alpha,\beta$. 
Using the Lorentz force result (\ref{FL3D}) we find,
\be
\partial_\varphi h_{\alpha\beta}\, \partial_\theta \alpha + \partial_\varphi h_{\beta\alpha}\, \partial_\theta \beta
= \partial_\theta h_{\alpha\beta} \, \partial_\varphi \alpha + \partial_\theta h_{\beta\alpha}\, \partial_\varphi \beta.
\label{EPconstraint}
\ee
This is solved by 
\be
h_{\alpha\beta} = f(r) \cM (\alpha), \qquad h_{\beta\alpha} = \tilde{f}(r) \tilde{\cM} (\beta),
\ee
where all functions are arbitrary. Interestingly, this class of solutions also encompasses the solution  (\ref{3DGS})
of the 3-D `Grad-Shafranov' equation (\ref{3DGSeq}) we derived for barotropic matter. 
In other words, for a barotropic non-axisymmetric system, (\ref{3DGS}) simultaneously 
solves (\ref{3DGSeq}) and (\ref{EPconstraint}). On the other hand, the 3-D stratified system is only constrained
by (\ref{EPconstraint}) and allows a broader family of solutions.

%%%%%%%%%%%%%%%%%%%%%%%%%%%%%%

\subsubsection{Case study: tilted torus magnetic field}
\label{sec:tilted}
To the best of our knowledge, there is only one analytic, non-axisymmetric magnetic field equilibrium in the literature: 
the so-called tilted torus magnetic fields \citep{lasky13b}.  These fields are physically motivated toy models for fields 
in proto-neutron stars.  Differential rotation or $r$-mode instabilities could wind up a strong toroidal component in the stellar core, 
whose axis of symmetry is aligned with the rotation axis of the star.  There is no a priori reason for the progenitor's field 
to be aligned with the remnant's rotation axis, implying the star is likely to have two misaligned components.  
Tilted-torus models posit that this misaligned field is a `purely poloidal' field, whose axis of symmetry is misaligned with 
the rotation axis, and hence misaligned with the axis of symmetry of the toroidal field.  While this field is strictly poloidal 
in some inclined frame (see below), it has a non-zero azimuthal contribution when expressed in the frame whose basis 
is aligned with the rotation axis of the star.  \citet{lasky13b} also showed how tilted torus-like fields arise from GRMHD simulations 
of proto-neutron stars.

Mathematically, tilted torus fields are described using two coordinate systems, ${\bf x}$ and ${\bf\bar x}$, where 
the barred coordinates are rotated by an angle $\xi$ 
in the $x$-$z$ plane with respect to the unbarred coordinates. The poloidal component is expressed in the usual way in the 
barred frame as 
\be
{\bf B}_\rP= \bar{\bnabla}\psi \times\bar{\bnabla}\bar{\varphi},
\ee
where $\bar{\bnabla}$ is the gradient operator in the barred coordinates, $\bar{r}=r$ 
and $\psi(r,\bar{\theta})=f(r)\sin^2\bar{\theta}$ is the stream function.  One has considerable freedom in 
choosing the radial function $f(r)$ \citep[e.g., see][for similar axisymmetric fields]{mastrano11,akgun13,mastrano13}, but 
herein we leave this function arbitrary. 

The toroidal component of the field is described in the usual way in the unbarred frame
\be
{\bf B}_\rT= \cT (\psi (r,\theta)) \bnabla\varphi.
\ee
In \cite{lasky13b}, a specific choice of $\cT (\psi)$ was made to satisfy integrability conditions for the force balance equations.  
We show below an equivalent description, but keep $\cT(\psi)$ general for the remainder of the derivation.

The total field is the sum of the toroidal and poloidal components, which is expressed in the unbarred frame as
\begin{align}
\bB= & \frac{2f}{r^2} \left(\cos\xi\cos\theta-\sin\xi\sin\theta\cos\varphi\right) \mathbf{\hat{r}}
             \notag\\
		& -\frac{f^\prime}{r}\left(\cos\xi\sin\theta+\sin\xi\cos\theta\cos\phi\right) \boldsymbol{\hat{\theta}}
		\notag\\
		& \frac{1}{r} \left ( f^\prime \sin\xi\sin\varphi +  \frac{\cT}{\sin\theta} \right) \boldsymbol{\hat{\varphi}},
\end{align}
where as always a prime denotes a derivative with respect to the argument. 

Putting this magnetic field through Eqn.~(\ref{Fconstrain1}) we derive a condition for the magnetic field
\be
f^2f^\prime  \cT^{\prime\prime}  \sin\xi\sin^2\theta\cos\theta\cos\varphi= 0.
\label{TTconstraint}
\ee
In general, there are only two conditions in which this equation is generally true.  
Firstly, when $\xi=0$, which represents the axisymmetric case, and hence equation~(\ref{Fconstrain1}) is trivially satisfied.  
In the non-axisymmetric case, one therefore has a condition on the toroidal field function, $\cT^{\prime\prime} = d^2 \cT/d^2\psi=0$.  
In \cite{lasky13b}, an integrability condition was derived that implied $\cT= c_1( \psi- c_0)$ for $\psi \ge c_0$ (where $c_0,c_1$ constants) 
and zero elsewhere.
This functional form of $\cT$ also satisfies the constraint equation (\ref{TTconstraint}).

%%%%%%%%%%%%%%%%%%%%%%%%%%%%%%%%%%%%%%%

\subsection{Digression: magnetic equilibria in the crust}
\label{sec:crust}

Real neutron stars are not just balls of fluid, but also have elastic crusts with magnetic fields 
threading both the core, crust and exterior region of the star.  In this section, we show the 
effect of an elastic force on possible magnetic equilibria in the crust.

From an EOS perspective, the crust behaves as a barotropic layer; however one should also
account for its elasticity. A crust threaded by a magnetic field is, in general, in a 
strained state described by a displacement field $\boldsymbol{\xi}$ and the crustal shear modulus $\mu$.

MHD equilibria in this strained crust are then described by the Euler equation\footnote{Realistic neutron star 
crusts consist of a solid lattice and a neutron superfluid (in their inner layers) and are therefore multifluid systems. 
Apart from a trivial density rescaling $\rho \to \rho_\rc$ in Eqn. (\ref{GScrust}), where $\rho_\rc$ is the density of 
the lattice, this property does not alter the conclusions of this section. Moreover, in this discussion we ignore
the crust's outermost thin fluid layer (ocean).},
\be
\bnabla (\delta h + \delta \Phi) = \frac{1}{\rho} \left (\, \bF_\rL + \bF_{\rm el} \, \right ).
\label{crust1}
\ee
Here, the Hookean force $\bF_{\rm el} = \mu \nabla^2 \boldsymbol{\xi} $ arises due to the crust elasticity 
(for simplicity we have assumed a uniform $\mu$ and an incompressible displacement $\bnabla \cdot \boldsymbol{\xi} = 0$, see
\citet{mcdermott88} for the full expression of $\bF_{\rm el}$). 
We thus have,
\be
\bnabla \times \left [\, \frac{1}{\rho} \left (\, \bF_\rL + \bF_{\rm el} \, \right ) \, \right ]= 0.
\label{GScrust}
\ee
It does not require much meditation on Eqn.~(\ref{GScrust})  to realise that the elastic force term 
adds an extra degree of freedom with which to balance the magnetic field.  One is therefore free to 
prescribe any magnetic field in the neutron star crust. This freedom is readily exploited in calculations of 
magnetic field evolution (due to Ohmic dissipation and the Hall effect) in neutron star crusts
\citep[see for instance,][]{pons07b,vigano13,gourgouliatos14}.

%%%%%%%%%%%%%%%%%%%%%%%%%%%%%%

\section{Multifluid neutron stars}
\label{sec:2fluid}

\subsection{Cold superfluid $npe$ matter }
\label{sec:coldsf}

The simplest model for superfluid neutron stars assumes $npe$ matter and consists of two 
fluids, namely, the neutron superfluid and the combined proton-electron conglomerate. 
The charged particles can be counted as one fluid because they are nearly comoving 
under the action of the magnetic forces. Their combined equation of motion is further simplified by the 
negligible electron mass and overall charge neutrality. It is also expected that protons are paired and 
form a type II superconductor at a very early stage of a neutron star's life. 
Superconductivity modifies the magnetic force -- this is now dominated by the tension of the quantised 
fluxtubes that thread the superconductor -- and almost eliminates any relative proton-electron motion. 
To a first approximation the thermal physics of the system can be ignored since the temperature of a 
neutron star falls well below its Fermi temperature ($T_\rF \sim 10^{12}\,\mbox{K}$) very soon after formation.  

The MHD equilibrium is now described by a pair of Euler equations \citep[see e.g.][]{glampedakis12a}
\begin{align}
&\rho_\rn \bnabla ( \mut_\rn + \Phi) = \bF_\rn,
\\
& \rho_\rp \bnabla ( \mut_\rpe + \Phi) = \bF_\rp,
\end{align}
where $\mut_\rn = \mu_\rn/m,~ \mut_\rpe = \mu_\rpe/m \equiv (\mu_\rp + \mu_\re )/m$ are chemical potentials per unit mass 
(we use a common baryonic mass $m=m_\rn = m_\rp \gg m_\re$) and $\mathbf{F}_\rn, \mathbf{F}_\rp$ are 
magnetic forces. Somewhat counterintuitively, there can be a magnetic force acting on the neutrons. This force 
arises when the protons are superconducting and is a consequence of the coupling between the proton fluxtubes 
and the neutron superfluid \citep[see][for details]{glampedakis11}.

For the total pressure we have the thermodynamical relation~\citep[e.g.,][]{prix04}
\be
\bnabla p = n_\rn \bnabla \mu_\rn + n_\rp \bnabla \mu_\rp + n_\re \bnabla \mu_\re 
=   \rho_\rn \bnabla \mut_\rn + \rho_\rp \bnabla \mut_\rpe,
\label{pres1}
\ee
where $n_\rn, n_\rp, n_\re$ are particle densities (with $n_\rp = n_\re$ as dictated by charge neutrality).

For the non-magnetic background equilibrium we have:
\be
\rho_\rn \bnabla ( \mut_\rn + \Phi )= \rho_\rp \bnabla ( \mut_\rpe + \Phi )= 0. 
\ee
These lead to  $\mut_\rpe = \mut_\rn \Rightarrow \mu_\rp + \mu_\re = \mu_\rn$ which is the condition for beta equilibrium. 
By adding the two equation we get the familiar equation for hydrostatic equilibrium  $\bnabla p = -\rho \bnabla \Phi$ with 
$\rho = \rho_\rp + \rho_\rn$  the total density.

The equations for the perturbed magnetic system are
\be
\rho_\rn \bnabla (\delta \mut_\rn + \delta \Phi)  = \bF_\rn,
\qquad 
\rho_\rp \bnabla (\delta \mut_\rpe + \delta \Phi)  = \bF_\rp.
\ee 
From (\ref{pres1}) we have  for the perturbed pressure,
\be
\delta p = \rho_\rn \delta \mut_\rn + \rho_\rp \delta \mut_\rpe
~\Rightarrow ~
\delta h = (1- \xp) \delta \mut_\rn + \xp \delta \mut_\rpe,
\ee
where $\xp = \rho_\rp/\rho$ is the proton fraction.

The two Euler equations can be combined in a natural way and produce an equivalent pair of a `total' 
and `difference' equations:
\begin{align}
& \bnabla (\delta h + \delta \Phi) - \delta \beta \bnabla \xp = \frac{1}{\rho} \left ( \bF_\rp + \bF_\rn \right ),
\label{eulersum1}
\\
& \rho  \bnabla \delta \beta =  \frac{1}{\xp}  \bF_\rp -  \frac{1}{1-\xp}  \bF_\rn. 
\label{eulerdif1}
\end{align}
The parameter 
\be
\delta \beta \equiv \delta \mut_\rpe -\delta \mut_\rn  
= \frac{1}{m} (\delta \mu_\rp + \delta \mu_\re -\delta \mu_\rn),
\ee
represents the departure from chemical equilibrium.

In order to proceed we need to specify the magnetic forces.  
For the case of normal (unpaired) protons there is only the `classical'  Lorentz force acting on the 
charged particles and we can set:
\be
\mathbf{F}_\rn =0, \qquad \mathbf{F}_\rp = \bF_\rL.
\ee
We note that this scenario could be the relevant one in (at least some) magnetars
when the interior field exceeds the threshold ($\sim 10^{15}-10^{16}\,\mbox{G} $) 
 for the suppression of superconductivity \citep[see][for details]{glampedakis11}.   

The difference Euler equation (\ref{eulerdif1}) becomes,
\be
\bnabla \delta \beta =  \frac{1}{\rho \xp}  \bF_\rL.
\ee
This obviously leads to a Grad-Shafranov equation:
\be
\bnabla \times \left (  \frac{1}{\xp \rho}  \bF_\rL  \right ) = 0.
\ee
Apart from a rescaling $\rho \to  \xp \rho $, this equation is identical to the one 
discussed earlier for single-fluid barotropic systems.

When protons are superconducting, the magnetic forces are given by \citep[e.g.,][]{glampedakis12a}:
\bear
&&\mathbf{F}_\rp = \frac{1}{4\pi} [\, \bnabla \times (H_\rc \hat{\bB} ) \, ] \times \bB  
-\frac{\rho_\rp}{4\pi} \bnabla \left ( B \frac{\partial H_\rc}{\partial \rho_\rp} \right ),
\label{Fp}
\\
\nonumber \\
&& \bF_\rn = -\frac{\rho_\rn}{4\pi} \bnabla \left ( B \frac{\partial H_\rc}{\partial \rho_\rn} \right ),
\label{Fn}
\eear
where $  \hat{\bB} = \bB/B$ and $H_\rc (\rho_\rp,\rho_\rn)$ is the lower critical field for
type II superconductivity \citep{tinkham96}.
It is easy to show that the gradient terms in (\ref{Fp}) and (\ref{Fn}) can be absorbed into 
the chemical potentials. This rearrangement amounts to a redefinition of chemical potentials:
\be
\delta \theta_\rn \equiv \delta \mut_\rn + \frac{B}{4\pi} \frac{\partial H_\rc}{\partial \rho_\rn},
\qquad 
\delta \theta_\rp \equiv \delta \mut_\rpe + \frac{B}{4\pi} \frac{\partial H_\rc}{\partial \rho_\rp}.
\ee
We similarly define,
\begin{align}
& \delta \tilde{\beta}  \equiv \delta \theta_\rp - \delta \theta_\rn = \delta \beta 
+ \frac{B}{4\pi} \left ( \frac{\partial H_\rc}{\partial \rho_\rp} - \frac{\partial H_\rc}{\partial \rho_\rn} \right ),
\\
& \delta \tilde{h} \equiv (1-\xp) \delta \theta_\rn + \xp \delta \theta_\rp.
\end{align}
In terms of the new parameters the Euler equations become:
\begin{align}
& \bnabla ( \delta \tilde{h} + \delta \Phi) - \delta \tilde{\beta} \bnabla \xp = \frac{1}{\rho} \bF_\rH,
\label{eulersum2}
\\
& \bnabla \delta \tilde{\beta} =  \frac{1}{\xp \rho}  \mathbf{F}_\rH, 
\label{eulerdif2}
\end{align}
where
\be
 \mathbf{F}_\rH = \frac{1}{4\pi} [\, \bnabla \times (H_\rc \hat{\bB} ) \, ] \times \bB, 
 \label{FH}
\ee
can be thought as the `Lorentz' part of the superconducting force. Thus the superconducting system 
too admits a Grad-Shafranov-type equation, i.e. 
\be
\bnabla \times \left (  \frac{1}{ \xp \rho}  \bF_\rH  \right ) = 0.
\ee
The upshot of this discussion is that, despite the non-barotropic nature of superfluid neutron stars made of 
cold $npe$ matter, a magnetic field in equilibrium cannot be freely prescribed. In a sense, the non-barotropic 
degree of freedom is undone by the second fluid degree of freedom and as a result the $\bB$ field has to solve 
a Grad-Shafranov equation, much alike barotropic systems. 
At a qualitative level the situation bears some resemblance to that of $g$-mode oscillations: 
a stratified  single-fluid system possesses a family of composition $g$-modes \citep{reisenegger92} -- these modes
disappear when the same system acquires a second fluid component \citep{prix02}.

%%%%%%%%%%%%%%%%%%%%%%%%%%%%%%%%%%%%%%%%%%%%%%%%%%%%%%%

\subsection{Hot superfluid $npe\mu$ matter}
\label{sec:hotsf}

In this section, while we continue considering superfluid/superconducting matter, we also take into account 
two more properties of realistic neutron stars: the unavoidable appearance of muons (as predicted by realistic EOSs)
and the finite temperature/entropy of matter. The resulting `hot'  $npe\mu$ model is the most realistic one used to date 
in the context of MHD equilibria in neutron stars. 

The muons are expected to appear above a density threshold representative of the outer core. Once present, they 
participate both in the beta reactions and charge neutrality of matter. The latter property is always preserved in the 
MHD approximation and therefore
\be
n_\rp = n_\re + n_\mu,
\label{neutral}
\ee
is always true. For a system in beta equilibrium (as is the case for the non-magnetic background star) we should have
\be
\mu_\rn = \mu_\rp + \mu_\re, \qquad  \mu_\rn = \mu_\rp + \mu_\mu \quad \Rightarrow \quad  \mu_\re = \mu_\mu.
\label{beta}
\ee
The addition of the magnetic field induces small deviations from this equilibrium and therefore we expect 
$\delta \mu_\rpe \neq \delta \mu_\rn$ and $ \delta \mu_\re \neq \delta \mu_\mu$.  
At the level of hydrodynamics the muons essentially behave as `heavy electrons' and are incorporated in the 
charged fluids conglomerate. In other words, they do not need a separate Euler equation. 

In the canonical multifluid framework used in this paper \citep[see e.g.,][]{prix04} entropy is viewed as one more fluid 
with velocity $\mathbf{v}_\rs$, chemical potential $\mu_\rs = T$ and number density $n_\rs = s$ where $s$ is the 
entropy density. 
In this language, for instance, heat conduction translates to a velocity lag between 
the entropy fluid and the other fluids.  For our purposes it makes sense to ignore conduction and assume that entropy is 
carried by  the normal particles (electrons/muons), so that $s=s_\re + s_\mu$. 
From the point of view of hydrodynamics this means that entropy and temperature should appear in the proton-electron-muon 
Euler equation.

The real importance of having a finite temperature in a neutron star core can be understood if we recall that the neutron/proton 
pairing energy is a bell-shaped function of $T$ which means that,  in certain regions, it may be exceeded 
by the thermal energy $k_{\rm B}T$, thus leading to a local suppression of neutron superfluidity (and to a lesser extent of proton 
superconductivity). 
As a result, the core may consist of adjacent multifluid and single fluid layers (the charged particle species are counted as a single fluid) 
of non-barotropic matter. Given that we have already discussed MHD equilibria in single-fluid stratified matter
(Section~\ref{sec:stratified}), here we focus on the finite-$T$ multifluid regions (we should note, however, that the boundary 
physics between the aforementioned layers is poorly understood and well beyond the scope of this paper -- our analysis may not 
apply in these boundaries).

The magnetic equilibrium in $npe\mu$ superfluid neutron stars is described, as before,  by a pair of Euler equations.
The neutron Euler equation is,
\be
\rho_\rn \bnabla ( \mut_\rn + \Phi) = \mathbf{F}_\rn ~ \Rightarrow ~ 
\rho_\rn \bnabla (\delta \mut_\rn + \delta \Phi ) = \bF_\rn.
\ee
The proton Euler contains most of the `new' physics:
\be
\rho_\rp \bnabla ( \mut_\rpe + \Phi) + \rho_\rp \frac{x_\mu}{x_\rp} \bnabla (\mut_\mu -\mut_\re) + s\bnabla T 
= \bF_\rp,
\label{eulerp1}
\ee
where we have used (\ref{neutral}) and assumed $m \gg m_\mu, m_\re$. Also, we have defined  the 
muon fraction $x_\mu = n_\mu/n$, where $n$ is the total particle number density.
The background part of (\ref{eulerp1}) leads to the beta equilibrium (\ref{beta}) in combination with a uniform
temperature $T$. For the perturbed part, we have:
\be
\rho_\rp \bnabla (\delta \mut_\rpe + \delta \Phi )  + \rho_\rp \frac{x_\mu}{\xp} \bnabla \delta \gamma 
+ s \bnabla \delta T = \bF_\rp,
\ee 
where we have defined the electron-muon chemical difference
\be
\delta \gamma \equiv \delta \mut_\mu -\delta \mut_\re = \frac{1}{m} \left (\, \delta \mu_\mu - \delta \mu_\re \, \right ).
\ee
The total pressure is given by,
\be
\bnabla p = \rho_\rn \bnabla \mut_\rn + \rho_\rp \bnabla \mut_\rpe + m n_\mu \bnabla \mut_\mu +  s\bnabla T, 
\label{pres2}
\ee
and this leads to
\be
\delta h =   \delta \mut_\rn + \xp \delta \beta + x_\mu \delta \gamma +  \tilde{s} \delta T, 
\label{dheq2}
\ee
where $ \tilde{s} = s /\rho $ is the specific entropy.

Following the procedure of the previous section we obtain the equivalent set of equations,
\begin{align}
& \bnabla ( \delta h + \delta \Phi) -  \delta \beta \bnabla \xp -\delta \gamma \bnabla x_\mu - \delta T \bnabla \tilde{s} 
=  \frac{1}{\rho} \left ( \bF_\rp + \bF_\rn \right ),  
\label{hotsum1}
\\
&\bnabla \left  ( \delta \beta + \frac{x_\mu}{\xp} \delta \gamma + \frac{\tilde{s}}{\xp}  \delta T \right ) 
- \delta \gamma \bnabla \left ( \frac{x_\mu}{\xp}\right )   
- \delta T \bnabla \left ( \frac{\tilde{s}}{\xp} \right )  
\nonumber \\
& =  \frac{1}{\rho_\rp}  \bF_\rp - \frac{1}{\rho_\rn}  \bF_\rn.  
\label{hotdiff1}
\end{align}
The key new terms in these equations are the gradients of $x_\mu$ and $\tilde{s}$ and their ratios with $x_\rp$.
Both terms are expected to be non-zero in realistic neutron stars and, between them, the muon composition gradient 
is expected to be the dominant effect (whenever muons are present) since the entropy term should scale 
with $T/T_{\rm F} \ll 1$. In fact, the muon composition terms can be as large as the proton/neutron terms in the 
Euler equations (the relative magnitude of the muon-entropy terms can also be inferred from the $g$-mode 
calculations of \citet{gusakov13} and \citet{passamonti16}).

In principle, the superconducting magnetic forces will also be modified to some extent.
For instance, it is well known that $H_\rc$ becomes a function of the temperature \citep[see][]{tinkham96}. 
The variational derivation of these forces \citep{glampedakis11} suggests that in addition to the 
$\partial H_\rc/\partial \rho_{\rn,\rp}$ gradient terms we should also expect the presence of a similar  
$\partial H_\rc/\partial T$ term.
At the same time, given that the basic structure of the fluxtube array remains the same irrespective of 
the presence of muons and temperature, we expect that $ \mathbf{F}_\rp,  \mathbf{F}_\rn$ will be given 
by expressions functionally similar to the ones of the previous section (in particular the $\mathbf{F}_\rH$ 
part of the force should retain its form, Eqn.~(\ref{FH})).

In fact, for the point we wish to make here, we do not need to specify the exact form of these forces: 
as evident from Eqn.~(\ref{hotdiff1}), the $ \delta \gamma \bnabla ( x_\mu/\xp ) $ and 
$ \delta T \bnabla \left ( \tilde{s}/\xp \right )$ terms prevent the magnetic forces from being equal to a total gradient. 
In other words, we can conclude that the realistic hot superfluid $npe\mu$ model does not admit a 
Grad-Shafranov-type equation. To what extent the system admits an arbitrary magnetic field is decided
by the underlying symmetry. 

%%%%%%%%%%%%%%%%%%%%%%

\subsubsection{Axisymmetric equilibria}
\label{sec:hotsfaxi}
Assuming an axisymmetric system we can show that, as was the case in the earlier single fluid non-barotropic model, 
the available equilibrium equations allow for an \textit{arbitrarily} prescribed magnetic field and magnetic forces 
(the latter can only depend on $\bB$, $H_\rc$ and background fluid parameters). 

The model at hand has eight degrees of freedom\footnote{Without counting the perturbed potential $\delta \Phi$ which 
can be readily calculated from the Poisson equation once $\delta \rho=\delta \rho_\rn + \delta\rho_\rp$ is known.}
associated with the perturbed fluid, namely,
$\{\delta h, \delta \beta, \delta \gamma, \delta T\}$ and $\{ \delta s, \delta n_\rn, \delta n_\rp,\delta \mut_\rn \}$.
There are also eight equations available at our disposal: four of them come from the Euler equations (\ref{hotsum1}), (\ref{hotdiff1}); 
three are `equations of state'  with a symbolic form 
$\delta n_x = f_x (\delta \beta,\delta \gamma, \delta \mut_\rn, \delta T), ~x=\{n,p,s\}$;
finally, the pressure equation~(\ref{dheq2}) can be written as 
$\delta \mut_\rn = f(\delta \beta,\delta \gamma, \delta T, \delta h) $.

We can thus see that once the Euler equations have been solved with respect to the subset 
$\{ \delta h, \delta \beta, \delta \gamma, \delta T \}$ 
the remaining functions can be readily obtained algebraically. From the $\theta$-Euler equations we obtain
\begin{align}
\delta h &= \zeta + \cS_\rp + \cS_\rn, 
\label{dh_sol1}
\\
\delta \beta &= \tilde{\zeta} - \frac{x_\mu}{x_\rp} \delta \gamma - \frac{\tilde{s}}{x_\rp} \delta T + \cS_\rp -\cS_\rn,
\label{dbeta_sol1}
\end{align}
where $\zeta(r)$ and $\tilde{\zeta}(r)$ are arbitrary functions and
\be
\cS_\rx (r,\theta) \equiv \frac{r}{\rho_\rx} \int d\theta F_\rx^\theta, \quad \rx=\{\rp,\rn\}.
\ee
Inserting these in the radial Euler components, we obtain an algebraic system for the remaining two
functions:
\begin{align}
\left (\frac{x_\mu}{x_\rp} \right )^\prime \delta \gamma + \left (\frac{\tilde{s}}{x_\rp} \right )^\prime 
\delta T  &= \tilde{\zeta}^\prime + \partial_r ( \cS_\rp - \cS_\rn )
\nonumber \\
& -\frac{1}{\rho_\rp} F^r_\rp + \frac{1}{\rho_\rn} F^r_\rn,
\\
\left (\frac{x_\mu}{x_\rp} \right )^\prime \delta \gamma + \left (\frac{\tilde{s}}{x_\rp} \right )^\prime 
\delta T& = \frac{1}{x_\rp} \left [\,  \zeta^\prime -x^\prime_\rp \tilde{\zeta}  + \partial_r (x_\rn \cS_\rn)  \right.
\nonumber \\
& \left. +  x_\rp^\prime \cS_\rn  + x_\rp \partial_r \cS_\rp  -\frac{1}{\rho} \left ( F^r_\rp + F^r_\rn \right ) \, \right ].
\label{system1}
\end{align}
The identical left-hand-sides mean that the system does not lead
to a unique solution for $\delta \gamma$ and $\delta T$. Moreover, from the equality of 
the right-hand-sides we get a differential relation between the two gauge functions:
\begin{align}
\zeta^\prime - ( x_\rp \tilde{\zeta} )^\prime &= \frac{1}{\rho} \left ( 1 + \frac{\rho_\rp}{\rho_\rn} \right ) F^r_\rn 
- \partial_r [ (x_\rp + x_\rn) \cS_\rn ]
\\
& = \frac{1}{4\pi} ( x_\rp + x_\rn)^\prime B \frac{\partial H_c}{\partial \rho_\rn} = 0.
\end{align}
The last line of this equation was obtained using (\ref{Fn}) and $x_\rp + x_\rn =1$
(note that the same result would hold for the case of normal protons, i.e. when $\bF_\rn=0$).

The upshot of this discussion is that, in an axisymmetric  neutron star composed of finite temperature 
$npe\mu$ matter, there is enough freedom in the perturbed fluid parameters to balance arbitrarily 
specified magnetic forces. Qualitatively speaking, the system behaves as the single-fluid non-barotropic 
star of Section~\ref{sec:stratified}. 
Interestingly, this is also true with respect to the $g$-mode oscillations (see related comment at the end 
of Section~\ref{sec:coldsf}): the presence of entropy and a muon component leads to the re-emergence of 
thermal/composition $g$-modes in superfluid neutron stars \citep{gusakov13,kantor14,passamonti16}.

%%%%%%%%%%%%%%%%%%%%%%%%%%%

\subsubsection{Non-axisymmetric equilibria}
\label{sec:hotsfnonaxi}
Moving on to general non-axisymmetric equilibria, we now need to take into account
the presence of two additional Euler equations (the $\varphi$-components). 
With six available equations for the four functions $\{ \delta h, \delta \beta, \delta \gamma, \delta T \}$
we expect to have two constraints for the magnetic force components. Indeed, following the same 
procedure as in Section~\ref{sec:nonaxi}, we can combine the $\theta$ and $\varphi$ components to obtain:
\begin{align}
\partial_\varphi ( F_\rp^\theta + F_\rn^\theta) &= \partial_\theta \left[\sin\theta 
\left( F^\varphi_\rp + F^\varphi_\rn\right)\right], 
\\
\partial_\varphi \left ( F_\rp^\theta -\frac{\rho_\rp}{\rho_\rn} F_\rn^\theta \right) &= 
\partial_\theta \left [\sin\theta 
\left (F^\varphi_\rp -\frac{\rho_\rp}{\rho_\rn} F^\varphi_\rn \right ) \right ].
\end{align}
From these we can produce the simpler relations:
\be
\partial_\varphi (F_\rx^\theta ) = \partial_\theta (\sin\theta F^\varphi_\rx), 
\qquad \rx = \{\rp,\rn\}.
\ee
Using the superconducting forces (\ref{Fp}) and (\ref{Fn}) we find that the gradient terms 
are mutually cancelled out, leaving just one non-trivial constraint equation:
\be
\partial_\varphi (F_\rH^\theta ) = \partial_\theta (\sin\theta F^\varphi_\rH).
\ee
This result is clearly very similar to the one found in Section~\ref{sec:nonaxi} and implies
that the  non-axisymmetric $npe\mu$ system does not admit an arbitrary magnetic field
equilibrium. We note that for non-superconducting protons we have 
$\bF_\rH \to \bF_\rL$ and we recover exactly our earlier result (\ref{Fconstrain1}). 

Finally, for the fluid variables we obtain the same algebraic equations (\ref{dh_sol1}), (\ref{dbeta_sol1}) 
and (\ref{system1}) as in the axisymmetric case (with $\zeta$ and $\tilde{\zeta}$ now functions of $r$ and $\varphi$).

%%%%%%%%%%%%%%%%%%%%%%%%%%%%%%%%%%%%%%%%

\section{General relativistic equilibria}
\label{sec:GR}

What we have learned so far about MHD equilibria in Newtonian stars 
can be carried over into the more realistic case of general relativistic stars. 
This section provides a  ``proof of principle''  analysis and as such it will suffice to
consider the simple case of a single-fluid axisymmetric GRMHD system. We show that 
(i) a barotropic EOS allows the formulation of a relativistic Grad-Shafranov equation
and (ii) once more realistic models of matter are considered (with stratification and departure from
chemical equilibrium), the Grad-Shafranov equation is lost and one is free 
to specify an arbitrary magnetic field equilibrium. 

The stress-energy tensor for a perfect fluid coupled with an electromagnetic (EM) field is,
\be
T^{\mu\nu} = T^{\mu\nu}_\rF + T^{\mu\nu}_\rEM,
\ee
where
\be
T^{\mu\nu}_\rF = ( \epsilon + p ) u^\mu u^\nu + p g^{\mu\nu},
\label{TF}
\ee
and
\be
T^{\mu\nu}_\rEM = \frac{1}{4\pi} \left(g_{\kappa\lambda} F^{\mu\kappa} F^{\nu\lambda} 
-\frac{1}{4} g^{\mu\nu} F_{\kappa\lambda} F^{\kappa\lambda} \right).
\label{TEM}
\ee
We have used standard notation, with $\epsilon$ denoting the energy density and 
$u^\mu$ the local four-velocity of fluid elements. The (antisymmetric) Faraday tensor $F_{\mu\nu}$ can 
be parametrised in terms of the four-potential $A^\mu$,
\be
F_{\mu\nu} = \nabla_\mu A_\nu - \nabla_\nu A_\mu =  \partial_\mu A_\nu - \partial_\nu A_\mu.
\ee
As usual, the introduction of $A_\mu$ comes with the gauge freedom bonus, $A_\mu \to A_\mu + \nabla_\mu f$.

The Maxwell equations for the EM field are,
\be
\nabla_\nu F^{\mu\nu} = 4\pi J^\mu, \qquad 
\nabla_{[\alpha} F_{\beta\gamma]} = 0,
\label{Max1}
\ee
where $J^\mu$ is the current density. 
An equivalent (and occasionally more practical) expression for the first equation is,
\be
\partial_{\nu} \left ( \sqrt{-g} F^{\mu\nu} \right ) = 4\pi \sqrt{-g} J^\mu,
\label{Max3}
\ee
where $g$ is the metric determinant.
The Lorentz force is defined as,
\be
F_\rL^\mu = - \nabla_\nu T_{\rEM}^{\mu\nu} = F^{\mu\nu} J_\nu,
\label{FL1}
\ee
where Eqns.~(\ref{Max1}) were used in the last step. 

The full GRMHD equations of motion are given by $\nabla_\nu T^{\mu\nu} = 0$. 
Upon projecting orthogonally with respect to $u^\mu$, we obtain the Euler equation:
\be
(\epsilon + p) u^\beta \nabla_\beta u_\alpha + \partial_\alpha p + u_\alpha u^\beta \partial_\beta p 
= F_{\rL \alpha}.
\label{euler1}
\ee
In addition, the MHD approximation is defined by the condition of a vanishing electric field,
\be
E_\mu = F_{\mu\nu} u^\nu = 0. 
\label{mhd}
\ee

%%%%%%%%%%%%%%%%%%%%%%%%%%%

\subsection{The relativistic Grad-Shafranov equation}
\label{sec:GS}

In order to study magnetic equilibria in GRMHD we first assume
a static and axisymmetric system. This implies a fluid four-velocity 
$u^\mu = ( u^t, 0,0,0)$ and a diagonal Weyl-type metric, 
\be
ds^2= g_{tt} dt^2 + g_{rr} dr^2 + g_{\theta\theta} d\theta^2 + g_{\varphi\varphi} d\varphi^2,
\label{weyl}
\ee
with $g_{\mu\nu} = g_{\mu\nu} (r,\theta)$. One can treat the EM field as a perturbation
of a spherical background star, implying the  metric 
$g_{\mu\nu} = \mbox{diag} [\, -e^{\nu(r)}, e^{\lambda(r)}, r^2, r^2\sin^2\theta\,]$ at leading order
with respect to the field \citep[see e.g. ][]{ciolfi09}.  However, the following derivation also applies using the metric~(\ref{weyl}).

Given these assumptions, the Euler equation (\ref{euler1}) becomes,
\be
F_{\rL\alpha} = \frac{1}{2} (\epsilon +p) g^{tt} \partial_\alpha g_{tt}  + \partial_\alpha p.
\label{euler2}
\ee
From this it follows that the azimuthal force vanishes, i.e. $F_{\rL \varphi} =0$.
Then, the $\varphi$-component of the Lorentz force law (\ref{FL1}) leads to
\be
J^r \partial_r \Psi  + J^\theta \partial_\theta  \Psi  = 0,
\label{Psi_eq1}
\ee
where we have defined the stream function\footnote{From the MHD condition~(\ref{mhd}) we have
$F_{\mu t} u^t = 0 \Rightarrow A_t = 0 $. Moreover, we can choose a gauge to make $A_\theta=0$.
The resulting EM potential is $A_\mu = (0,A_r,0,\Psi)$.} $\Psi (r,\theta) \equiv A_\varphi$.

The components $J^r, J^\theta$ can be calculated from (\ref{Max3}):
\be
J^i =  \frac{1}{4\pi \sqrt{-g}} \partial_j \left ( \sqrt{-g} F^{ij} \right ),
\label{Jeq}
\ee
where $i,j = \{r,\theta\}$ and $i\neq j$. Using this result in (\ref{Psi_eq1}),
\be
\partial_r \Psi \partial_\theta \left ( \sqrt{-g} F^{r\theta} \right ) 
- \partial_\theta \Psi \partial_r \left ( \sqrt{-g} F^{r\theta} \right ) = 0.
\ee
This implies
\be
\sqrt{-g} F^{r\theta} = T (\Psi),
\label{Teq1}
\ee
with $T$ an arbitrary function representing the toroidal field degree of freedom.

From the remaining Maxwell equation components, after using $F_{t\alpha} =0$ and 
$F_{\varphi\alpha} = - \partial_\alpha \Psi $, we obtain $J^t=0$ and 
\be
J^\varphi = -\frac{1}{4\pi \sqrt{-g}} \partial_j \left (\, \frac{\sqrt{-g}}{g_{\varphi\varphi} g_{jj}} \partial_j \Psi \, \right ),
\label{Jph2}
\ee
where summation over $j$ is assumed.  We are now in a position to calculate the Lorentz force in terms of the stream functions. 
We find the trivial results
$F_{\rL t} = F_{\rL \varphi} = 0$ and
\be
F_{\rL i} = \left [\, J^\varphi + \frac{ T T^\prime }{4\pi g_{tt} g_{\varphi\varphi}}  \, \right ] \partial_i \Psi
\equiv \cA_{\rm GR} \, \partial_i \Psi,
\label{FL3}
\ee
where $T^\prime = dT/d\Psi$. Using (\ref{Jph2}),
\be
\cA_{\rm GR} = -\frac{1}{4\pi }\left [\,   \frac{\partial^2_j \Psi}{g_{\varphi\varphi} g_{jj}}    
+ \partial_j \left (\, \frac{\sqrt{-g}}{g_{\varphi\varphi} g_{jj}} \,\right ) \frac{\partial_j \Psi}{\sqrt{-g}}
- \frac{T T^\prime}{g_{tt} g_{\varphi\varphi} } \, \right ].
\ee
The last step of this analysis is the manipulation of the Euler equation (\ref{euler2}). 
This, however, entails making a choice for the EOS of matter.

%%%%%%%%%%%%%%%%%%
\subsection{Barotropic matter}
\label{sec:baroGS}
For a barotropic EOS we have relations $p=p(\epsilon)$ and $\epsilon=\epsilon(n)$
where $n$ is the total baryon number density. From the first thermodynamical law we can obtain the 
Euler relation
\be
p + \epsilon = \mu n, \qquad \mu \equiv \frac{d\epsilon}{dn},
\label{eos1}
\ee
where $\mu$ is the chemical potential. These can be written in a differential form,
\be
\partial_\alpha \epsilon =  \mu   \partial_\alpha n = \frac{p+\epsilon}{n}  \partial_\alpha n.
\label{eos2}
\ee
For the pressure gradient we obtain,
\be
\partial_\alpha p =  -\partial_\alpha \epsilon + \partial_\alpha (\mu n) 
= (\epsilon+p)  \frac{\partial_\alpha \mu}{\mu}.
\label{dp_baro}
\ee
Inserting this in the Euler equation, 
\be
F_{\rL \alpha} = (\epsilon + p) \partial_\alpha \left (\, \frac{1}{2} \nu + \log \mu \, \right )  
\equiv (\epsilon + p) \partial_\alpha X,
\label{euler3}
\ee
where we have used $g_{tt} = -e^\nu$.

After equating (\ref{euler3}) and (\ref{FL3}) we have,
\be
 \cA_{\rm GR} \, \partial_i \Psi = (\epsilon + p) \partial_i X.
\ee
This implies that $X=X(\Psi)$ and
\be
\cA_{\rm GR} = (\epsilon + p) M(\Psi)
\ee
with $M$ arbitrary. Written explicitly, this result is:
\begin{align}
& \partial^2_j \Psi  + \partial_j \log \left (\, \frac{\sqrt{-g}}{g_{\varphi\varphi} g_{jj}} \,\right )  \partial_j \Psi
- \frac{ g_{jj}  }{g_{tt}  } T T^\prime =
\nonumber \\
&= -4\pi  (\epsilon + p)  g_{\varphi\varphi} g_{jj} M(\Psi),
\end{align}
with $ j = \{r,\theta\} $ summed. We have thus arrived at the desired relativistic Grad-Shafranov equation. 
This of course means that magnetic field equilibria in axisymmetric relativistic barotropic stars are not arbitrary. 
For actual calculations of relativistic Grad-Shafranov equilibria the reader can consult e.g. \citet{colaiuda08,ciolfi09}.

%%%%%%%%%%%%%%%%%%%%%%%%

\subsection{Non-barotropic matter: restoring freedom in MHD equilibrium}
\label{sec:noGS}

A more sophisticated model for matter should account for the presence of stratification 
and deviations from chemical beta equilibrium. As already pointed out in previous sections, 
these effects require a non-barotropic EOS. As a more realistic benchmark model, in this section we assume 
a multi-constituent, single-fluid system \citep[for a review see][]{andersson07b}. 
The constituents comprise neutrons, protons and electrons with number densities $n_\rx$, $\rx=\{\rn,\rp,\re\}$. 
Only two of these are independent since we always require local charge neutrality, i.e. $n_\rp = n_\re$. 

The upgraded EOS is of the form $\epsilon=\epsilon (n_\rx)$ and from this we have,
\be
d\epsilon = \sum_\rx  \mu_\rx dn_\rx =  \mu_\rn dn_\rn + \mu_{\rm pe} dn_\rp,
\ee
where the chemical potentials are defined as $\mu_\rx \equiv \partial \epsilon /\partial n_\rx$ and 
$\mu_{\rm pe} = \mu_\rp + \mu_\re$. We can write a similar expression with covariant/partial
derivatives:
\be
\partial_\alpha \epsilon =  \mu_\rn \partial_\alpha n_\rn 
+ \mu_{\rm pe} \partial_\alpha n_\rp.
\ee
In terms of $n=n_\rn + n_\rp$ and the proton fraction $x_\rp = n_\rp/n$ this becomes,
\be
\partial_\alpha \epsilon = \mu_\rn \partial_\alpha n + \beta \partial_\alpha (n x_\rp),
\label{de1}
\ee
where  $\beta \equiv \mu_{\rm pe} - \mu_\rn$ (not to be confused with the
slightly different $\beta$ parameter of the Newtonian models).

The total pressure of the system is given by,
\be
p =  -\epsilon + \sum_\rx n_\rx  \mu_\rx = \epsilon + n \mu_\rn + n x_\rp \beta.
\ee
Taking the derivative of this and using (\ref{de1}) we find, 
\begin{align}
\partial_\alpha p &= n \left [\, \partial_\alpha (\mu_\rn + x_\rp \beta) - \beta \partial_\alpha x_\rp \, \right ]
\notag \\
&= \frac{\epsilon+p}{\mu_\rn + x_\rp \beta} \left [\, \partial_\alpha (\mu_\rn + x_\rp \beta) 
- \beta \partial_\alpha x_\rp \, \right ].
\end{align}
This result reduces to the barotropic expression (\ref{dp_baro}) for (i) a system in 
beta equilibrium, $\beta=0$,  or (ii) a uniform composition, $\partial_\alpha x_\rp =0$ 
(in which case $\mu =\mu_\rn + x_\rp \beta $).

Our earlier result (\ref{FL3}) for the Lorentz force is valid irrespective of the EOS.
However, the same is not true for the Euler equation (\ref{euler3}). For the present non-barotropic
model that equation is replaced by
\be
F_{\rL \alpha} = (\epsilon + p) \left (\, \partial_\alpha \tilde{X} - \beta \partial_\alpha x_\rp \, \right ),
\ee
where $\tilde{X} \equiv \nu/2 + \log ( \mu_\rn + x_\rp\beta)$.
Therefore, 
\be
\frac{\cA_{\rm GR}}{\epsilon+p} \partial_i \Psi = \partial_i \tilde{X} -\beta \partial_i x_\rp 
~ \Rightarrow ~
\frac{\cA_{\rm GR}}{\epsilon+p} \partial_i \Psi \neq M(\Psi).
\ee
In other words, there is \textit{no} Grad-Shafranov equation. 
It does not take much more work to show that for any arbitrary magnetic field $\{ \Psi, T(\Psi)\}$ 
the available fluid degrees of freedom can be chosen so that the equilibrium equations are satisfied. 
We have thus arrived at the same conclusion as in the case of Newtonian stars (Section~\ref{sec:stratified}).

%%%%%%%%%%%%%%%%%%%%%%%%%%%%%%%

\section{Concluding remarks}
\label{sec:conclusions}

The allowed space of MHD equilibria in neutron stars is dependent on the nature of matter, namely the equation of state.  
In this paper, we have surveyed one's freedom to arbitrarily prescribe magnetic equilibria for different types of matter
(i.e. number of distinct fluids, composition and entropy gradients, deviations from chemical equilibrium), 
degrees of symmetry (i.e. axisymmetry/non-axisymmetry) and types of gravity (i.e. Newtonian/general relativistic). 
This freedom depends on whether there are available fluid degrees of freedom to balance the magnetic force. 
Our results are summarised as follows (see Table \ref{tab:overview} for a bird's eye view summary):

\begin{enumerate}

\item Axisymmetric systems have a rich spectrum of arbitrariness with respect to MHD equilibria.
We have found that the usual Grad-Shafranov equation is not only a property of simple barotropic 
stellar models. It can also control the MHD equilibrium in stratified matter provided the latter is multifluid 
(e.g. $npe$ matter with neutron superfluidity). However, the addition of muons and entropy (hot $npe\mu$ matter)
nullifies the Grad-Shafranov equation, eventually leading to freely specifiable magnetic fields 
(as in the case of single-fluid non-barotropic systems). 
Among other things, this freedom implies an arbitrary relative strength between the poloidal and 
toroidal field components. 

\item In non-axisymmetric systems, the additional azimuthal components of the equations of motion 
prevent the magnetic field from being arbitrarily specified. The resulting constraint, at least for the case
of non-superconducting matter, leads to a pair of Grad-Shafranov-like equations for the magnetic field's 
scalar degrees of freedom (i.e. the Euler potentials). 

\item The transition from Newtonian to general relativistic gravity does not alter the above conclusions 
(but increases the complexity of the various equilibrium equations). 

\end{enumerate}

This paper has solely focused on MHD equilibria. The dynamical \textit{stability} of these equilibria
is a completely different  and much harder question, with obvious repercussions for their astrophysical
relevance. Recent work \citep{lander12} suggests that barotropic equilibria are 
generically unstable, but this may not be true for more realistic non-barotropic systems since the buoyancy force
emerging in stratified matter is known to enhance stability~\citep[e.g.,][]{akgun13}. Another avenue for instability could
be provided by the interplay between rotation and magnetic field crust-core coupling during the initial spin down 
of newly formed neutron stars~\citep{glampedakis15}.  

Somewhat surprisingly, our work has some bearing on the nature of the Hall equilibrium in neutron star crusts.
This equilibrium refers to the asymptotic $t \to +\infty$ state of the magnetic induction equation when the field is
sourced by electron currents (this is the so-called electron-MHD) and is set to evolve due to the Hall term, i.e.
$ \partial_t \bB = \bnabla \times (\mathbf{v}_\re \times \bB ) \propto \bnabla \times \{ (\mathbf{J}_\re \times \bB)/n_\re \}$.
We can immediately deduce that, in axisymmetry, the condition for Hall equilibrium is identical to Eqn.~(\ref{curl1}),
thus leading to the Grad-Shafranov equation (\ref{GradS}) with $\rho \to \rho_\re$~\citep{gourgouliatos14}.
It is straightforward to generalise this result to the full non-axisymmetric case; we predict that the 
Hall equilibrium should be described by our 3-D Grad-Shafranov equation~(\ref{3DGSeq}). 
It will be interesting to test this prediction with the recently developed  numerical framework for 3-D Hall evolution
in neutron star crusts~\citep{gourgouliatos16}.

We hope that this paper will serve as a stepping stone for modelling the next-generation MHD equilibria in
realistic neutron stars.

%%%%%%%%%%%%%%%%%%%%%%%%%%%%%%%%%%%%%%%%%%%%%%%
\section*{acknowledgments}
KG is supported by the Ram\'{o}n y Cajal Programme of the Spanish Ministerio de 
Ciencia e Innovaci\'{o}n and by NewCompstar (a COST-funded Research Networking Programme). 
PDL is supported by an Australian Research Council Discovery Project DP1410102578. 
The authors would like to thank Ashley Bransgrove for useful discussions.

%%%%%%%%%%%%%%%%%%%%%%%%%
\bibliographystyle{mn2e}
\bibliography{MHD_bib}

\begin{thebibliography}{}

\bibitem[\protect\citeauthoryear{{Akg{\"u}n}, {Reisenegger}, {Mastrano} \&
  {Marchant}}{{Akg{\"u}n} et~al.}{2013}]{akgun13}
{Akg{\"u}n} T.,  {Reisenegger} A.,  {Mastrano} A.,    {Marchant} P.,  2013,
  Mon. Not. R. Astron. Soc., 433, 2445

\bibitem[\protect\citeauthoryear{{Andersson} \& {Comer}}{{Andersson} \&
  {Comer}}{2007}]{andersson07b}
{Andersson} N.,  {Comer} G.~L.,  2007, Living Reviews in Relativity, 10, 1

\bibitem[\protect\citeauthoryear{Braithwaite}{Braithwaite}{2007}]{braithwaite07}
Braithwaite J.,  2007, A\&A, 469, 275

\bibitem[\protect\citeauthoryear{Braithwaite}{Braithwaite}{2009}]{braithwaite09}
Braithwaite J.,  2009, Mon. Not. R. Astron. Soc., 397, 763

\bibitem[\protect\citeauthoryear{{Bucciantini}, {Pili} \& {Del
  Zanna}}{{Bucciantini} et~al.}{2015}]{bucciantini15}
{Bucciantini} N.,  {Pili} A.~G.,    {Del Zanna} L.,  2015, Mon. Not. R. Astron.
  Soc., 447, 3278

\bibitem[\protect\citeauthoryear{Ciolfi, Ferrari, Gualtieri \& Pons}{Ciolfi
  et~al.}{2009}]{ciolfi09}
Ciolfi R.,  Ferrari V.,  Gualtieri L.,    Pons J.~A.,  2009, Mon. Not. R.
  Astron. Soc., 397, 913

\bibitem[\protect\citeauthoryear{Ciolfi \& Rezzolla}{Ciolfi \&
  Rezzolla}{2012}]{ciolfi12}
Ciolfi R.,  Rezzolla L.,  2012, Astrophys. J., 760, 1

\bibitem[\protect\citeauthoryear{{Ciolfi} \& {Rezzolla}}{{Ciolfi} \&
  {Rezzolla}}{2013}]{ciolfi13}
{Ciolfi} R.,  {Rezzolla} L.,  2013, Mon. Not. R. Astron. Soc., 435, L43

\bibitem[\protect\citeauthoryear{Colaiuda, Ferrari, Gualtieri \& Pons}{Colaiuda
  et~al.}{2008}]{colaiuda08}
Colaiuda A.,  Ferrari V.,  Gualtieri L.,    Pons J.~A.,  2008, Mon. Not. R.
  Astron. Soc., 385, 2080

\bibitem[\protect\citeauthoryear{{Fujisawa} \& {Eriguchi}}{{Fujisawa} \&
  {Eriguchi}}{2013}]{fujisawa13}
{Fujisawa} K.,  {Eriguchi} Y.,  2013, Mon. Not. R. Astron. Soc, 432, 1245

\bibitem[\protect\citeauthoryear{Glampedakis, Andersson \& Lander}{Glampedakis
  et~al.}{2012}]{glampedakis12a}
Glampedakis K.,  Andersson N.,    Lander S.~K.,  2012, Mon. Not. R. Astron.
  Soc., 420, 1263

\bibitem[\protect\citeauthoryear{Glampedakis, Andersson \&
  Samuelsson}{Glampedakis et~al.}{2011}]{glampedakis11}
Glampedakis K.,  Andersson N.,    Samuelsson L.,  2011, Mon. Not. R. Astron.
  Soc., 410, 805

\bibitem[\protect\citeauthoryear{{Glampedakis} \& {Lasky}}{{Glampedakis} \&
  {Lasky}}{2015}]{glampedakis15}
{Glampedakis} K.,  {Lasky} P.~D.,  2015, Mon. Not. R. Astron. Soc., 450, 1638

\bibitem[\protect\citeauthoryear{{Gourgouliatos} \& {Cumming}}{{Gourgouliatos}
  \& {Cumming}}{2014}]{gourgouliatos14}
{Gourgouliatos} K.~N.,  {Cumming} A.,  2014, Phys. Rev. Lett., 112, 171101

\bibitem[\protect\citeauthoryear{{Gourgouliatos}, {Wood} \&
  {Hollerbach}}{{Gourgouliatos} et~al.}{2016}]{gourgouliatos16}
{Gourgouliatos} K.~N.,  {Wood} T.~S.,    {Hollerbach} R.,  2016, Proceedings of
  the National Academy of Science, 113, 3944

\bibitem[\protect\citeauthoryear{{Gusakov} \& {Kantor}}{{Gusakov} \&
  {Kantor}}{2013}]{gusakov13}
{Gusakov} M.~E.,  {Kantor} E.~M.,  2013, Mon. Not. R. Astron. Soc., 428, L26

\bibitem[\protect\citeauthoryear{Haberl}{Haberl}{2007}]{haberl07}
Haberl F.,  2007, Astrophys. Space Sci., 308, 181

\bibitem[\protect\citeauthoryear{Haskell, Samuelsson, Glampedakis \&
  Andersson}{Haskell et~al.}{2008}]{haskell08}
Haskell B.,  Samuelsson L.,  Glampedakis K.,    Andersson N.,  2008, Mon. Not.
  R. Astron. Soc., 385, 531

\bibitem[\protect\citeauthoryear{{Kantor} \& {Gusakov}}{{Kantor} \&
  {Gusakov}}{2014}]{kantor14}
{Kantor} E.~M.,  {Gusakov} M.~E.,  2014, Mon. Not. R. Astron. Soc., 442, L90

\bibitem[\protect\citeauthoryear{Kaspi}{Kaspi}{2010}]{kaspi10}
Kaspi V.~M.,  2010, PNAS, 107, 7147

\bibitem[\protect\citeauthoryear{Kiuchi, Yoshida \& Shibata}{Kiuchi
  et~al.}{2011}]{kiuchi11}
Kiuchi K.,  Yoshida S.,    Shibata M.,  2011, A\&A, 532, 17

\bibitem[\protect\citeauthoryear{Lander}{Lander}{2013}]{lander12b}
Lander S.~K.,  2013, Phys. Rev. Lett., 110, 071101

\bibitem[\protect\citeauthoryear{Lander \& Jones}{Lander \&
  Jones}{2012}]{lander12}
Lander S.~K.,  Jones D.~I.,  2012, Mon. Not. R. Astron. Soc., 424, 482

\bibitem[\protect\citeauthoryear{Lasky \& Melatos}{Lasky \&
  Melatos}{2013}]{lasky13b}
Lasky P.~D.,  Melatos A.,  2013, Phys. Rev. D, 88, 103005

\bibitem[\protect\citeauthoryear{Lasky, Zink, Kokkotas \& Glampedakis}{Lasky
  et~al.}{2011}]{lasky11}
Lasky P.~D.,  Zink B.,  Kokkotas K.~D.,    Glampedakis K.,  2011, Astrophys.
  J., 735, L20

\bibitem[\protect\citeauthoryear{McDermott, {van Horn} \& Hansen}{McDermott
  et~al.}{1988}]{mcdermott88}
McDermott P.~N.,  {van Horn} H.~M.,    Hansen C.~J.,  1988, Astrophys. J., 325,
  725

\bibitem[\protect\citeauthoryear{{Mastrano}, {Lasky} \& {Melatos}}{{Mastrano}
  et~al.}{2013}]{mastrano13}
{Mastrano} A.,  {Lasky} P.~D.,    {Melatos} A.,  2013, Mon. Not. R. Astron.
  Soc., 434, 1658

\bibitem[\protect\citeauthoryear{Mastrano, Melatos, Reissenegger \&
  Akg\"un}{Mastrano et~al.}{2011}]{mastrano11}
Mastrano A.,  Melatos A.,  Reissenegger A.,    Akg\"un T.,  2011, Mon. Not. R.
  Astron. Soc., 417, 2288

\bibitem[\protect\citeauthoryear{{Mereghetti}, {Pons} \&
  {Melatos}}{{Mereghetti} et~al.}{2015}]{mereghetti15}
{Mereghetti} S.,  {Pons} J.~A.,    {Melatos} A.,  2015, Space Science Reviews,
  191, 315

\bibitem[\protect\citeauthoryear{{Palapanidis}, {Stergioulas} \&
  {Lander}}{{Palapanidis} et~al.}{2015}]{palapanidis15}
{Palapanidis} K.,  {Stergioulas} N.,    {Lander} S.~K.,  2015, Mon. Not. R.
  Astron. Soc., 452, 3246

\bibitem[\protect\citeauthoryear{{Passamonti}, {Andersson} \&
  {Ho}}{{Passamonti} et~al.}{2016}]{passamonti16}
{Passamonti} A.,  {Andersson} N.,    {Ho} W.~C.~G.,  2016, Mon. Not. R. Astron.
  Soc., 455, 1489

\bibitem[\protect\citeauthoryear{{Pons} \& {Geppert}}{{Pons} \&
  {Geppert}}{2007}]{pons07b}
{Pons} J.~A.,  {Geppert} U.,  2007, A\&A, 470, 303

\bibitem[\protect\citeauthoryear{{Prix}}{{Prix}}{2004}]{prix04}
{Prix} R.,  2004, Phys. Rev. D, 69, 043001

\bibitem[\protect\citeauthoryear{{Prix} \& {Rieutord}}{{Prix} \&
  {Rieutord}}{2002}]{prix02}
{Prix} R.,  {Rieutord} M.,  2002, A\&A, 393, 949

\bibitem[\protect\citeauthoryear{Reisenegger}{Reisenegger}{2009}]{reisenegger09}
Reisenegger A.,  2009, A\&A, 499, 557

\bibitem[\protect\citeauthoryear{Reisenegger \& Goldreich}{Reisenegger \&
  Goldreich}{1992}]{reisenegger92}
Reisenegger A.,  Goldreich P.,  1992, Astrophys. J., 395, 240

\bibitem[\protect\citeauthoryear{Tinkham}{Tinkham}{1996}]{tinkham96}
Tinkham M.,  1996, Introduction to superconductivity.
McGraw-Hill, N.Y

\bibitem[\protect\citeauthoryear{{Vigan{\`o}}, {Rea}, {Pons}, {Perna},
  {Aguilera} \& {Miralles}}{{Vigan{\`o}} et~al.}{2013}]{vigano13}
{Vigan{\`o}} D.,  {Rea} N.,  {Pons} J.~A.,  {Perna} R.,  {Aguilera} D.~N.,
  {Miralles} J.~A.,  2013, Mon. Not. R. Astron. Soc., 434, 123

\bibitem[\protect\citeauthoryear{{Woods} \& {Thompson}}{{Woods} \&
  {Thompson}}{2006}]{woods06}
{Woods} P.~M.,  {Thompson} C.,  2006, {Soft gamma repeaters and anomalous X-ray
  pulsars: magnetar candidates}.
pp 547--586

\bibitem[\protect\citeauthoryear{{Yahalom} \& {Lynden-Bell}}{{Yahalom} \&
  {Lynden-Bell}}{2008}]{yahalom08}
{Yahalom} A.,  {Lynden-Bell} D.,  2008, Journal of Fluid Mechanics, 607, 235

\end{thebibliography}

\end{document}